# The Model of Fast Contraction Flows for Polymeric Melts

JAE-HYEUK JEONG *and* ARKADY I. LEONOV

*Department of Polymer Engineering*
*The University of Akron*
*Akron, Ohio 44325-0301*

**Abstract**
The present paper develops an isothermal model for fast (high Deborah number) contraction flows of polymers from a reservoir to a die of circular or rectangular cross-sections. Two components of composite flow models in different regions of the flow domain are connected in succession with the use of asymptotic matching boundary conditions. These components are inhomogeneous elongation and a modified unsteady shearing. It is constructed based on a stable and descriptive set of viscoelastic constitutive equations with the specified material parameters. The present model demonstrates a reasonable good agreement to the experimental/direct numerical data and improves for the higher values of Deborah number and contraction ratio cases without any adjustable parameters. Also it is applicable to any constitutive equation and performs with easy PC numerical calculations.

## INTRODUCTION

The present paper develops an approximate analysis for fast isothermal *contraction* flows of polymer melts. The complex polymer flows occur in the axi-symmetric composite channel shown in Fig.1. It consists of the reservoir (a channel with larger cross-sectional area), connected to the die (a channel with smaller cross-sectional area). The cross-sectional geometries of reservoir and die are considered to be either circular or rectangular, with high ratios of the cross-sectional areas. These geometries are widely used in several high-speed polymer processing operations, such as extrusion and injection molding, as well as in testing of polymer fluids using slit/capillary rheometers. This type of flow also represents a benchmark problem for numerical studies of polymer flows at high Deborah (*De*) numbers. The complexity of these analyses, especially for high *De* number flows, results from nonlinear viscoelastic character of constitutive equations (CE's) for polymeric liquids and the effect of sudden change in the geometry of the composite channels.

It is convenient to schematically subdivide the entire complex flow domain in five characteristic geometric regions: (i) the reservoir pre-entrance, (ii) reservoir entrance, (iii) the die entrance, (iv) the die land region, and (v) the die exit region. Additionally there is the free surface extrudate swell flow.
We assume that a *developed* simple shearing flow of Poiseuille type occurs in the region (i) located far away upstream from the die entrance. In the reservoir region (ii), which is close to the die entrance, the previous Poiseuille flow dramatically changes to a converging type of flow suited for entering the polymer stream in the die. In this region, many experiments for polymer melts and all they direct numerical simulations observe a secondary flow (vortices) at the reservoir entrance corners. In the *developing flow* in the die entrance region (iii), additional restructuring happens from the upstream converging flow back to a simple shearing flow in the die. The developing flow is accompanied by the release/dissipation of the stored elastic energy accumulated in the previous converging flow. If the die is long enough, the die land region (iv) may exist as a continuation of the developing flow region. Here the *developed* shearing flow of Poiseuille type occurs whose characteristics are independent of the longitudinal die coordinate. At the very end of the die exit region (v), the die flow restructures once again to the free-surface extrudate flow. Although this type of re-structuring is dynamically important, it happens in a relatively short region. Therefore flow in this region can be approximately treated as continuing previous (developing or developed) flow up to the die end with a sudden change in the flow type at the die end cross section.

Previously many experimental and computational works have been done for the flows of polymer melts and solutions in the flow regions (ii)-(iv). Flow visualization experiments, such as laser Doppler velocimetry (LDV) [1,2], particle tracer methods [3,4], laser speckle velocimetry [5,6] and birefringence methods [7], have been commonly employed to analyze the development of flow patterns near the die entrance.
The contraction type of polymer flow also attracted attention as one of the benchmark problems in computational viscoelastic fluid dynamics. It is because this type of flow involves the geometrical complexity along with non-trivial set of equations for description of momentum, continuity and nonlinear viscoelasticity of polymeric liquids. After the extensive review of present finite element method (FEM), the mixed FEM was proposed with the demonstration of advantages in simulating the planar contraction problem [8].

In the review of the pre-entrance/entrance contraction flow, White et al. [9] pointed out that the choice of CE's also influences the numerical stability and evolutionary nature of CE's. Some numerical studies [10,11] of fast viscoelastic flows near sharp entrance corners revealed non-convergence with mesh refinement using the upper convected Maxwell model (UCM). It was partially attributed to the occurrence of corner singularities. Notable research efforts [12,13] with different CE's and several benchmark problems, including the contraction flow, demonstrated that the Leonov type of CE is the most stable and the corner singularities do not affect computations for this CE up to the value of Deborah number approaching 700 [13].

Along with the instability problems, the authors of Ref.[9] emphasized the importance of extensional flow in entry flow. They revealed that the common failure of numerical prediction of entry flow comes from inabilities of most CE's to describe well both extensional and shear flows. The dominancy of extensional property in fast entrance flows of viscoelastic liquids has been discussed many times in experimental papers [7,14]. Cogswell [15] first made an effort to confirm his idea using his very simplistic empirical flow model. Although in our opinion, this model is of doubtful value, the Cogswell general idea seems to be extremely important and has been confirmed in many experimental papers cited above.

Starting from the paper [15], several attempts were made to develop a simplified empirical flow model for the entrance converging flow. The paper [16] tried to improve the Cogswell model using empirical equations for simple shearing and simple elongation. The papers [17,18] involved in computations a specific CE of simple integral type. They calculated the stress-strain rate relations in the "Protean" (stream function-longitudinal) coordinate system well known in the theory of convective diffusion [19]. In their formulation, they also employed an arbitrarily chosen "minimum principle" with empirical distribution of the longitudinal velocity. In spite of the model arbitrariness and unstable character of original simple integral CE, the paper reported successful agreement between the model calculations and experimental data.

The present work develops a model for high *De* number contraction flow with abrupt change in entrance geometry. It starts with a set of general, descriptive, and stable CE's and simplifies the flow complexity using two different characteristics of the flows in different geometrical regions. In the pre-reservoir region (i), the flow is modeled as developed simple shearing flow. Then following the Cogswell idea [15], the main converging stream of polymer liquid in the reservoir region (ii) is modeled as an inhomogeneous extensional flow similar to that employed in the fiber spinning. In the region (iii) where the entrance die flow is developing, we use once again the CE's for unsteady simple shearing, with evolution of elastic strains originated from the entrance reservoir flow. The time derivative is simplistically treated in this unsteady flow model as space convective derivative as has been done in many papers before (e.g. see Ref. [20]). Additionally, more detailed momentum balance equation is used in the region (iii) to stabilize numerical calculations. Finally, the characteristics of developed flow in the region (iv) are calculated as limiting regime of flow in very long tubes. The transition between flow types in these different flow regions is modeled by specific matching boundary conditions. We have also taken into account the viscoelastic drag affected the main extensional flow in the region (ii).

The presented flow model, designed to save computational cost, demonstrates a reasonable good agreement with existing experimental results and direct numerical simulations. Additionally, there are two remarkable feature of this approach. The first is that no single adjustable parameter is involved in the flow modeling. The second is that the model calculations demonstrate better comparison with experimental data or/and direct calculations when *De* number and contraction ratio are increased.

## 2. THEORY

### 2.1. Constitutive Equation

In the following, we use a multi-mode viscoelastic CE of differential type. It has been recently employed [21,22] for detailed and consistent description of all experimentally available rheological data for five polymeric melts, while satisfying all the stability constraints [23].

For each relaxation mode with relaxation time $\theta_k$ and elastic modulus $G_k$, the CE operates with a modal elastic Finger tensor $\underline{\underline{c}}_k$ whose evolution for incompressible case is described by the equation:

$$2\theta_k \stackrel{\nabla}{\underline{\underline{c}}}_k + b[\underline{\underline{c}}_k^2 + \underline{\underline{c}}_k (I_{2k} - I_{1k})/3 - \underline{\underline{\delta}}] = \underline{\underline{0}}, \quad \stackrel{\nabla}{\underline{\underline{c}}} \equiv \dot{\underline{\underline{c}}} - \underline{\underline{c}} \cdot \underline{\nabla}\mathbf{v} - (\underline{\nabla}\mathbf{v})^T \cdot \underline{\underline{c}}$$

$$I_{1k} = tr\underline{\underline{c}}_k, \quad I_{2k} = tr\underline{\underline{c}}_k^{-1}, \quad I_{3k} = \det \underline{\underline{c}}_k = 1; \quad b = b(I_1, I_2). \tag{1}$$

Here $\underline{\nabla}\mathbf{v}$ is the velocity gradient tensor, $\underline{\underline{\delta}}$ is the unit tensor, and *b* is a positive scaling factor for the relaxation time, which is generally a function of invariants, $I_{1k}$ and $I_{2k}$ with parameters independent of the mode number. This function has been specified for various polymeric melts and elastomers [21,22]. The extra stress tensor $\underline{\underline{\sigma}}_e$ and the total stress $\underline{\underline{\sigma}}$ are defined as:

$$\underline{\underline{\sigma}} = -p\underline{\underline{\delta}} + \underline{\underline{\sigma}}_e; \quad \underline{\underline{\sigma}}_e = \sum_k \underline{\underline{\sigma}}_k = 2\sum_k \underline{\underline{c}}_k \cdot \partial W_k / \partial \underline{\underline{c}}_k.$$

$$W_k \equiv G_k(T) w_k(I_{1k}) = \frac{3G_k(T)}{2(n+1)}[(I_{1k}/3)^{n+1} - 1]. \qquad (2)$$

Here $p$ is the isotropic pressure, $\underline{\underline{\delta}}$ is the unit tensor and $\underline{\underline{\sigma}}_k$ is the extra stress tensor for $k^{th}$ relaxation mode. Here we have simplified the general form of modal elastic potential $W_k$ proposed in [48]. The potential in (2) is represented via the Hookean modulus, $G_k(T)$, slightly depending on temperature, and the non-dimensional function, $w_k$, which is characterized by only one, mode-independent numerical parameter $n$. The general choice of the function $b(I_1, I_2)$ considered in Refs.[21,22] and parameter $n$ depends only on rheological behavior of polymer. The formulations and values of these non-linear terms are specified in the Table 1. Then the only rheological parameters to be specified are those that describe the linear relaxation spectrum, $\{G_k, \theta_k\}$.

## 2.2. Modeling the Entrance Contraction Flow

Our schematic interpretation of flow outlined in the Introduction, hypothesizes that two basic regions of polymer flow exist in the reservoir. When the reservoir is long enough, there is the flow region (i) where there is a developed steady shear flow of polymeric melts. We call the flow in this region *the far field entrance flow*. Additionally, there is the flow region (ii) near the die entrance where the main stream of flow converges to the die accompanying sometimes by weak entrance vortices at the flow periphery near the entrance corner. We call the flow in this region *the near field entrance flow*. Let $x_1$ be the axial coordinate directed along the symmetry axis in flow direction with the origin ($x_1 = 0$) at the die entrance. Then the far and near field flows are located in the respective axial domains ($x_1 < -l$) for the region (i) and ($0 > x_1 > -l$) for the region (ii); the coordinate, $x_1 = -l$, being the parameter searched for. Although the change of the flow characters from the far field to the near field entrance flow needs a small transitional axial length, saying $\delta$, we asymptotically will use the coordinate $x_1 = -l$ as the boundary where sudden change in flow type happens.

    We initially approximate the contraction flow in the near field entrance region (ii) as an inhomogeneous extensional (or "jet") flow that commonly occurs either in polymer fiber spinning or in polymer sheet processing operations. We assume that such an approach is approximately valid when the flow *De* number is high enough. In order to compare our calculations with the literature birefringence data obtained for low *De* number flows, we then develop more detailed approach involving into analysis the secondary shear flow, and include the viscoelastic drag from the near-wall secondary

flow in dynamics of the jet. We further demonstrate that for moderate *De* number flows, the contribution of the secondary flow might be important for initial development of the jet. The details are following "*Jet Flow with Drag from Secondary Viscoelastic Flows*" section. However this contribution is insignificant for flows with higher *De* numbers.

Two types of reservoir geometries are analyzed: the reservoir of the slit type (Fig.1a) with cross-sectional sizes $2L_R$ (thickness) and $L$ (width) where $L_R \ll L$, and the reservoir of circular geometry with the reservoir radius $R_R$. To avoid unnecessary duplications and simplify notations, a simultaneous analysis of flows in both types of geometries is provided below, with asterisked formulae attributed to the circular geometry. The same notations $\{x_1, x_2, x_3\}$ are used below for the Cartesian and cylindrical coordinate systems to describe the respective flows in the slit ($-L_R < x_2 < L_R$) and circular ($0 < x_2 < R_R$) geometries. The index 1 stands for axial ($z$), 2 – for transversal ($y$ or $r^*$), and 3 – for neutral ($x$ or $\varphi^*$) directions.

**Flow in the Far Field Entrance Region (i):** $\{x_1 < -l,\ L_R > x_2 > -L_R \text{ or } 0 < x_2 < R_R\}$.

This is a steady simple shear flow with the velocity gradient, $\underline{\nabla v}$, the extra stress, $\underline{\underline{\sigma}}_e$, and the elastic strain, $\underline{\underline{c}}_k$, tensors whose matrices are:

$$\underline{\nabla v} = \dot{\gamma}\begin{pmatrix} 0 & 0 & 0 \\ 1 & 0 & 0 \\ 0 & 0 & 0 \end{pmatrix},\ \underline{\underline{\sigma}}_e = \begin{pmatrix} \sigma_{11,e} & \sigma_{12,e} & 0 \\ \sigma_{12,e} & \sigma_{22,e} & 0 \\ 0 & 0 & \sigma_{33,e} \end{pmatrix},\ \underline{\underline{c}}_k = \begin{pmatrix} c_{11,k} & c_{12,k} & 0 \\ c_{12,k} & c_{22,k} & 0 \\ 0 & 0 & 1 \end{pmatrix}. \quad (3)$$

Substituting (3) into (1) written for steady state yields the known solution [44]:

$$c_{11,k} = \frac{\sqrt{2}\zeta_k}{\sqrt{\zeta_k + 1}},\ c_{22,k} = \frac{\sqrt{2}}{\sqrt{\zeta_k + 1}},\ c_{12,k} = \frac{\sqrt{\zeta_k^2 - 1}}{\zeta_k + 1},\ c_{33,k} = 1,\ \zeta_k(\dot{\gamma}) = \sqrt{1 + (2\theta_k \dot{\gamma}/b)^2}$$

$$\sigma_{ij,e} = \sum_k G_k (I_{1,k}/3)^n c_{ij,k};\quad \dot{\gamma} = dv_1/dx_2;\quad I_{1,k} = I_{2,k} = c_{11,k} + c_{22,k} + 1 = \sqrt{2(\zeta_k + 1)} + 1. \quad (4)$$

Here $\dot{\gamma}$ is the shear rate; $v_1$ is the only non-zero velocity component in simple shearing.

The momentum balance equations result in:

$$\sigma_{12,e}(x_2) = \frac{dP}{dx_1} \cdot x_2;\quad \sigma_{12,e}(x_2)^* = \frac{1}{2}\frac{dP}{dx_1} \cdot x_2. \quad (5, 5^*)$$

Here *P* is the pressure and $\sigma_{12,e}$ is the shear stress. The expression for flow rate *Q* is:

$$Q = 2L\int_0^{L_R} v_1 dx_2 = -2L\int_0^{L_R} x_2 \dot{\gamma} dx_2; \quad Q^* = 2\pi \int_0^{R_R} v_1 x_2 dx_2 = -\pi \int_0^{R_R} x_2^2 \dot{\gamma} dx_2. \quad (6, 6^*)$$

The flow rate in the steady flow $Q$ is constant and considered below as given. When $Q$ is specified, the shear rate, $\dot{\gamma}(x_2)$, can be found from Eqs.(4), (5) and (6). It enables us to compute the complete profiles of rheological variables for the steady shear flow up to the distance $l$ from the entrance.

We employed in simple numerical procedures the Newton-Rhapson method to find the gap-wise shear rate profile for a given value of flow rate $Q$, while satisfying Eqs. (4)-(6). The numerical integration of Eq.(6) was performed using the trapezoidal method. We should note that the location of the borderline between the far and the near field reservoir entrance flows has not been found yet.

**Flow in the Near Field Entrance Region (ii)** $\{0 > x_1 > -l, \; L_R > x_2 > -L_R \; (0 < x_2 < R_R)\}$.

*Jet Flow Approach*

We assume that when flow De number is high enough, the contraction flow in the near field entrance region (ii) can be modeled as an inhomogeneous extensional, "jet" flow with flat profiles of longitudinal velocity and extensional stress. This assumption is directly based on the Cogswell idea [15] and the flow visualization experiments [7], where the measured core stress distribution for high enough De number flows becomes more jet-like near the die. This simplified rough modeling ignores vortices at the entrance corners and neglects the shear effects near the reservoir walls.

We now consider the jet flow in the slit and circular reservoir geometries. In the slit case the jet flow can be approximated as an inhomogeneous planar elongation flow. In the circular case, we use the inhomogeneous simple elongation approximation. Then the corresponding expressions for the matrices of elastic strain tensors, $\underline{\underline{c}}_k$, velocity gradient tensor, $\underline{\nabla v}$, and the extra stress tensor, $\underline{\underline{\sigma}}_e$, in the jet are:

$$\underline{\underline{c}}_k = \begin{pmatrix} \lambda_k^2 & 0 & 0 \\ 0 & \lambda_k^{-2} & 0 \\ 0 & 0 & 1 \end{pmatrix}, \; \underline{\nabla v} = \dot{\varepsilon}\begin{pmatrix} 1 & 0 & 0 \\ 0 & -1 & 0 \\ 0 & 0 & 0 \end{pmatrix}, \; \underline{\underline{\sigma}}_e = \begin{pmatrix} \sigma_{11} & 0 & 0 \\ 0 & \sigma_{22} & 0 \\ 0 & 0 & \sigma_{33} \end{pmatrix} \quad (7)$$

$$\underline{\underline{c}}_k^* = \begin{pmatrix} \lambda_k^2 & 0 & 0 \\ 0 & \lambda_k^{-1} & 0 \\ 0 & 0 & \lambda_k^{-1} \end{pmatrix}, \; \underline{\nabla v}^* = \dot{\varepsilon}\begin{pmatrix} 1 & 0 & 0 \\ 0 & -1/2 & 0 \\ 0 & 0 & -1/2 \end{pmatrix}, \; \underline{\underline{\sigma}}_e^* = \begin{pmatrix} \sigma_{11} & 0 & 0 \\ 0 & \sigma_{22} & 0 \\ 0 & 0 & \sigma_{33} \end{pmatrix} \quad (7^*)$$

Here $\dot{\varepsilon}$ is the elongation rate. Inserting (7) into (1) and employing the "convective approximation", $d/dt \approx \overline{v}_1 d/dx_1$, commonly used for the description of these inhomogeneous elongation flows, yields:

$$\frac{1}{\lambda_k} \cdot \overline{v}_1 \frac{d\lambda_k}{dx_1} + \frac{b}{4\theta_k}(\lambda_k^2 - \frac{1}{\lambda_k^2}) = \dot{\varepsilon}; \quad \frac{1}{\lambda_k} \cdot \overline{v}_1 \frac{d\lambda_k}{dx_1} + \frac{b}{6\theta}(\lambda_k^2 - \frac{1}{\lambda_k})(1 + \frac{1}{\lambda_k}) = \dot{\varepsilon}. \tag{8,8*}$$

Here $\overline{v}_1$ is the axial velocity averaged over the jet cross-section, and $\dot{\varepsilon} = d\overline{v}_1/dx_1$ is the elongation rate. Similar to homogeneous planar and simple elongations, the expression for the elongation stress (a total axial stress in jet cross-sections), $\sigma_{ext}$, in the multi-mode case is presented as:

$$\sigma_{ext} = \sum_k G(I_{1,k}/3)^n \cdot (\lambda_k^2 - \lambda_k^{-2}), \quad I_{1k} = 1 + \lambda_k^2 + \lambda_k^{-2}; . \tag{9}$$

$$\sigma_{ext}^* = \sum_k G(I_{1,k}/3)^n \cdot (\lambda_k^2 - \lambda_k^{-1}), \quad I_{1k} = \lambda_k^2 + 2\lambda_k^{-1}. \tag{9*}$$

Due to the incompressibility condition held in the jet flow,

$$\overline{v}_1(x_1) = Q/A(x_1), \quad \dot{\varepsilon} \equiv \frac{d\overline{v}_1}{dx_1} = -\frac{Q}{A^2} \cdot \frac{dA}{dx_1}. \tag{10}$$

Here $A(x_1)$ is the cross-sectional area of jet, the flow rate $Q$ being the same as in the region (i). With Eq.(10), the evolution equations (8) for each relaxation modes take the form:

$$\frac{Q}{A\lambda_k} \cdot \frac{d\lambda_k}{dx_1} + \frac{b}{4\theta_k}(\lambda_k^2 - \frac{1}{\lambda_k^2}) = -\frac{Q}{A^2} \cdot \frac{dA}{dx_1}; \tag{11}$$

$$\frac{Q}{A\lambda_k} \cdot \frac{d\lambda_k}{dx_1} + \frac{b}{6\theta_k}(\lambda_k^2 - \frac{1}{\lambda_k})(1 + \frac{1}{\lambda_k}) = -\frac{Q}{A^2} \cdot \frac{dA}{dx_1}. \tag{11*}$$

The set of equations (11) for each mode is not closed yet for unknowns $A$ and $\lambda_k$. To make the closure, we involve another assumption that the force acting on the jet is constant: $F_{ext} = \sigma_{ext} \cdot A = const$, which is exactly the same as for the real inhomogeneous elongation flow. This condition is expected to work well enough for high *De* number flows where the jet flow dominates. Using this condition, Eq.(9) is written in the final, form:

$$F_{ext} \equiv A \cdot \sigma_{ext} = A \cdot \sum_k G(I_{1,k}/3)^n \cdot (\lambda_k^2 - \lambda_k^{-2}) = const, \quad I_{1k} = 1 + \lambda_k^2 + \lambda_k^{-2}; \tag{12}$$

$$F_{ext}^* \equiv A^* \cdot \sigma_{ext}^* = A \cdot \sum_k G_k(I_{1k}/3)^n \cdot (\lambda_k^2 - 1/\lambda_k) = const; \quad I_{1k}^* = \lambda_k^2 + 2/\lambda_k. \tag{12*}$$

Equations (11) and (12) represent a closed set. Appropriate boundary conditions for these equations are (see Fig 1):

$$x_1 = -l: \quad A = A_l = 2L \cdot L_R, \quad \lambda_k = \lambda_k^l; \quad A = A_l = \pi R_R^2, \quad \lambda_k = \lambda_k^l. \tag{13,13*}$$

$$x_1 = 0: \qquad A = A_0 = 2L \cdot L_D, \quad \lambda_k = \lambda_k^o.; \quad A = A_0 = \pi R_D^2, \quad \lambda_k = \lambda_k^0 \qquad (14, 14^*)$$

Here the known geometric parameters $L$ and $L_D$ ($R_D$) are respectively the width of slit die the half thickness (radius) of slit planar (circular) die. Note that as soon as the values of elongation elastic stretches $\lambda_k^l$ are known, the steady nonlinear boundary value problem is completely closed. To find the values $\lambda_k^l$ we need to use at the unknown boundary $x_1 = -l$, an additional matching condition (A4) derived in Appendix A.

### *Jet Flow with Drag from Secondary Viscoelastic Flows*

We now present a simplified analysis of the secondary shear flow effect in the modeling of "*jet flow*". Even if this effect is negligible when the value of *De* number is high enough, the birefringence data [7] obtained for moderate flow rates, clearly show that the drag from the secondary shear flow outside of the core jet creates an extra force affecting the jet development. It is uneasy to distinguish in this case the difference between the core jet (elongation) flow and near-wall secondary (shear) flow. To avoid this uncertainty and keep computational simplicity, we simply define the jet as the near-axial stream that completely contributes to the flow rate. It means that the secondary flow is circulatory and confined in a closed, wedge-shape domain near the corner. The circulatory flow is overwhelmingly observed in numerous experimental data for high *De* number polymer flows and its existence was well documented in all direct computations. Then including the viscoelastic drag from the near-wall secondary flow in dynamics of the jet, results in calculating the effect of secondary flows on the normal force and jet sizes. We will model the secondary shear flow using the lubrication layer approximation (LLA) [20], which is commonly applied to a wedge-like shearing between two well-defined solid boundaries. In our case the difficulty is that the jet profile is unknown and searched for.

Within the LLA approach, the momentum balance equations for the secondary flows are written in a simplified form:

$$\sigma_{12,e}(x_2) = x_2 dP/dx_1 + c_1; \quad \sigma_{12,e}(x_2)^* = (x_2/2)dP/dx_1 + c_1/x_2 \qquad (15, 15^*)$$

Here the pressure $P$ and unknown parameter $c_1$ are considered as independent of $x_2$, and $\sigma_{12,e}$ is the shear stress. In the circulatory flow, the flow rate is equal to zero, i.e.

$$Q_{\sec} = -2L \int_{L_{Ex}}^{L_R} x_2 \dot{\gamma} dx_2 = 0; \quad Q_{\sec}^* = -\pi \int_{R_{Ex}}^{R_R} x_2^2 \dot{\gamma} dx_2 = 0. \qquad (16, 16^*)$$

Here $L_{Ex}$ ($R_{Ex}$) is the semi-thickness/radius of jet and $L_R$ ($R_R$) is the semi-thickness/radius of slit/circular reservoir. The non-slip boundary conditions at the reservoir wall and the continuity of velocity at jet surface are:

$$x_2 = L_R \text{ or } R_R: \quad v_{1,\sec} = 0 \qquad (17)$$

$$x_2 = L_{Ex} \text{ or } R_{Ex}: \quad v_{1,\sec} = \overline{v}_{1,\text{jet}} = Q/A(x_1) \qquad (18)$$

Consider now the effect of drag forces on the jet flow. The local force balance between the extensional jet force and shear drag acting in the $x_1$ direction is:

$$\frac{d}{dx_1}(\sigma_{ext} \cdot L_{Ex}) = -\sigma_{12,surf}(x_1), \quad \frac{d}{dx_1}(\sigma^*_{ext} \cdot \pi R^2_{Ex}) = -2\pi R_{Ex} \cdot \sigma^*_{12,surf}(x_1) \qquad (19,19^*)$$

Here $\sigma_{ext}$ is the extensional stress acting on the jet and $\sigma_{12,surf}$ is the shear stress at the jet surface due to circulatory drag flow.

The finite difference iterative algorithm used for calculations in this case is more complicated than that for the pure jet approach. The brief description of the algorithm on the example of plain flow is as follows. Assuming that at a certain value of $x_1$ the jet profile, $L_{Ex}(x_1)$, and the extensional stress, $\sigma_{ext}(x_1)$, are known. It means that the jet velocity $\bar{v}_{1,jet}(x_1) = Q/A(x_1)$ is known too. Then the shearing, circulatory viscoelastic flow problem (15)-(18) is well defined and has been elaborated long ago in extrusion whenever the "flow curve", the function $\sigma_{12}(\dot{\gamma})$ assumed to be monotonously increasing is given. Then the inverse function, $\dot{\gamma} = \partial v_{1,sec}/\partial x_2 = -\psi(\sigma_{12})$, is well defined too. In our case we use in the computations the function $\sigma_{12,e}(\dot{\gamma})$ defined in (4). Taking into account Eqs.(15)-(18), the gap-wise distribution of the axial velocity and the condition for determining two unknown parameters, $dP/dx_1$ and $c_1$, are:

$$v_{1,sec} = \int_{x_2}^{L_R} \psi(x_2'dP/dx_1 + c_1) dx_2',$$

$$\bar{v}_{1,jet} = \int_{L_{Ex}}^{L_R} \psi(x_2 dP/dx_1 + c_1) dx_2; \quad \int_{L_{Ex}}^{L_R} x_2 \psi(x_2 dP/dx_1 + c_1) dx_2 = 0. \qquad (20)$$

Using a numerical procedure outlined below and the formulae (20), the problem of circulatory viscoelastic flow is solved for any value $x_1$, including the determination of the drag shear stress $\sigma_{12,surf}(x_1)$. Then the calculations of Eqs.(11) and (12) with variable extensional force $F_{ext}$ are performed using a small increment $\Delta x_1$ to obtain the values of the jet profile and extensional force at the level $x_1 + \Delta x_1$. It is evident that these calculations show the positive increase in the extensional force, $\Delta F_{ext}$, caused by the drag. The above iterative procedure started at the upstream corner, $x_1 = -l$, where $L_{Ex} \to L_R$ ($R_{Ex} \to R_R$). Unfortunately, at this point the problem for the secondary flow displays a singular stress behavior. To proceed with numerical computations we needed to reveal the character of this singularity. The analysis given in Appendix B shows that the singularity is integrable, meaning that the forces and energies are restricted in the singular point. Using this analysis, Appendix B also provides a brief description of our calculation effort near the singular point.

The numerical procedure for the unknown parameters $\{\dot{\gamma}(x_2), c_1\}$ is performed using the common matrix solver subroutine, with satisfying Eqs. (15, 15*), (16, 16*) and boundary conditions (17) and (18). These calculations completely solve the secondary flow problem, including the shear profile at a given axial position and the drag at the jet surface. After that the jet extensional force for the next step, $x_1 + \Delta x_1$, can be computed from Eq. (19, 19*). It is easy to understand that the circulatory shear flow model will decrease the jet length $l_1$. This is because the drag increases the extensional jet force, which in turn makes the jet thinner. Therefore when the initial (reservoir) and final (die) area are fixed the transition from the initial and final jet thickness is getting shorter.

**Matching Condition at the Boundary Between the two Entrance Flow Regions.**
The jet flow model has not yet been completed because the initial elastic stretches $\lambda_k^l$ ($k=1,2,...$) in Eq.(13, 13*) are unknown. To determine them we employ at the unknown boundary, $x_1 = -l$, an energetic matching condition (A4) derived in Appendix A for each $k^{th}$ nonlinear Maxwellian mode. This energetic condition, proposed *at hoc* first in Ref. [52], provides the modeling opportunity for asymptotic matching of various flow parameters (homogeneously or inhomogeneously distributed) in different regions of flow at an effective "interface" with discontinuities in the values of these parameters. In real flows, these parameters are certainly changed continuously from one asymptotic flow region to another. The description of these continuous changes is the subject of direct numerical simulations.

The matching condition (A4) takes the form:

$$x_1 = -l: \quad \frac{1}{A_l}\int_\Omega v_1^{sh} \cdot W_k^{sh} dA = v_1^{jet} \cdot W_k^{jet}. \tag{21}$$

Here $W_k^{sh}$ and $W_k^{jet}$ are elastic potentials for each $k^{th}$ nonlinear Maxwell mode in the far-field shearing and near-field jet flows, $v_1^{sh}$ and $v_1^{jet}$ are the axial velocities of shear and jet flow, and $A_R$ is the reservoir cross-sectional area. Using formulae (4) and (9) for simple shearing and elongation flows, reduce Eq.(21) for any $k^{th}$ relaxation mode to the form:

$$x_1 = -l: \quad \frac{L}{Q}\int_0^{L_R} v_1^{sh}(c_{11,k} + c_{22,k} + 1)^{n+1} dx_2 = [(\lambda_k^l)^2 + (\lambda_k^l)^{-2} + 1]^{n+1}, \quad (k=1,2,...); \tag{22}$$

$$x_1 = -l: \quad \frac{2\pi}{Q}\int_0^{R_R} x_2 \cdot v_1^{sh}(c_{11,k} + c_{22,k} + 1)^{n+1} dx_2 = [(\lambda_k^l)^2 + 2(\lambda_k^l)^{-1}]^{n+1}, \quad (k=1,2,...). \tag{22*}$$

Here we used the fact that at $x_1 = -l$, the jet cross sectional area coincides with that for the reservoir. Equation (22) allows to find the initial jet elastic stretches $\lambda_k^l$ from the known elastic strain tensor profile for far field shear flow, $c_{ij,k}$, which are found in Eqs. (4)-(6).

Now inserting the values of $\lambda_k^l$ in Eq. (9) with the reservoir cross sectional area, the total longitudinal jet force can be easily calculated at $x_1 = -l$. As the result, the further jet area depending on axial position, $A(x_1)$, can be expressed via $\sigma_{ext}$ through the calculated value of $F_{ext}$. Equations (11) and (12) along with conditions (13), (14) and (22) represent a simple model for steady entrance flow calculations. Mathematically, this problem is reduced to a well-defined boundary value problem with the parameter $l$ being the eigenvalue. The above model for reservoir flow also provides the calculations of developing flow in the die with appropriate boundary conditions.

We used in our calculations a numerical procedure with trapezoidal integration and the root finding routine for Eq.(22), along with the Newton-Raphson methods for solving Eq. (11) and (12) until the boundary condition (14) is satisfied.

When *De* number and the contraction ratio of the flow are not high enough to satisfy our prior assumptions for the jet flow model, the other set of force condition for the jet model with drag from secondary shear flow may work better. The calculations using this more complicated model have been demonstrated in the previous Section and in Appendix B.

Although the described jet approach oversimplifies the reservoir near field flow in the vicinity $x_1 = -l$, it is getting more realistic when the flow is closer to the entrance. It is also remarkable that this approach has no fitting parameters.

## 2.3. Modeling the Die Developing Flow: { $L_z > x_1 > 0$, $L_D > x_2 > -L_D$ $(0 < x_2 < R_D))$ }

We now model the developing flow in the die for both the slit and circular geometries. We will treat the developing flow as a version of non-steady viscoelastic Poiseuille (die) shearing flow where the time differentiating operator is substituted for the axial convective space operator as: $d/dt \approx \overline{v}_1 \partial / \partial x_1$. Here $\overline{v}_1 = Q/A_0$ is the die average longitudinal velocity, and $A_0$ is the cross-sectional die area equal to $2L \cdot L_D$ and $\pi (R_D)^2$ for the slit and circular die geometries, respectively. This approach is similar to that that was successfully used for the calculations of developing viscous flows in long tubes [20].

The *evolution equation* for each $k^{th}$ mode (the index $k$ is omitted) for both the die geometries, with the structure of elastic strain tensors $\underline{\underline{c}}_k$ shown in Eq.(3), has the form:

$$\begin{cases} \overline{v}_1 \frac{\partial c_{12}}{\partial x_1} + \frac{b}{2\theta} c_{12}(c_{11} + c_{22}) = c_{22} \frac{\partial v_1}{\partial x_2} & \text{("12" component)} \\ \overline{v}_1 \frac{\partial c_{22}}{\partial x_1} + \frac{b}{2\theta}(c_{12}^2 + c_{22}^2 - 1) = 0 & \text{("22" component)} \\ c_{11} c_{22} - c_{12}^2 = 1 & \text{incompressibility condition} \end{cases} \quad (23)$$

The shear and longitudinal normal stress components for the extra stress tensor are:

$$\sigma_{12} = \sum_k G_k (I/3)^n c_{12,k}; \quad \sigma_{11} = \sum_k G_k (I/3)^n c_{11,k}. \quad (24)$$

As the consequence of the formulae (23) and (24), we further use only the longitudinal component of the momentum balance equation,

$$\frac{\partial p}{\partial x_1} = \frac{\partial \sigma_{11}}{\partial x_1} + \frac{\partial \sigma_{12}}{\partial x_2}; \quad \frac{\partial p}{\partial x_1} = \frac{\partial \sigma_{11}}{\partial x_1} + \frac{1}{x_2}\frac{\partial (x_2 \sigma_{12})}{\partial x_2} \qquad (25, 25^*)$$

and the continuity equation:

$$\frac{\partial v_1}{\partial x_1} + \frac{\partial v_2}{\partial x_2} = 0; \quad \frac{\partial v_1}{\partial x_1} + \frac{1}{x_2}\frac{\partial (x_2 v_2)}{\partial x_2} = 0 \qquad (26, 26^*)$$

The condition for flow rate to be constant is:

$$Q = 2L\int_0^{L_D} v_1 dx_2 = -2L\int_0^{L_D} x_2 \dot{\gamma} dx_2 = const; \qquad (27)$$

$$Q = 2\pi \int_0^{R_D} x_2 v_1 dx_2 = -\pi \int_0^{R_D} x_2^2 \dot{\gamma} dx_2 = const. \qquad (27^*)$$

The non-slip boundary conditions for the components of velocity are:

$$x_2 = \pm L_D: \quad v_1 = v_2 = 0,; \quad x_2 = R_D: \quad v_1 = v_2 = 0. \qquad (28, 28^*)$$

In order to find the boundary conditions at the entrance, $x_1 = 0$, we use once again conditions (A4) from Appendix A for matching the jet flow in the reservoir and entrance shearing flow in the die. Two simplifying assumptions are used. We firstly assume that all the "initial" variables at the beginning ($x_1 = 0$) of the developing die flow are distributed homogeneously across the entrance cross-section. This is in accord with calculations [20] of developing viscous flows. We secondly assume that the shear stress at the entrance cross-section $x_1 = 0$ is absent. This follows from the well-known fact that after sudden imposition of shear velocity, there is no instant shear stress response. Using conditions (A4) with the above assumptions yields for any $k^{th}$ relaxation mode:

$$x_1 = 0: \quad v_1^0 = \overline{v}_1 \quad (v_2 = 0); \quad c_{12}^0 = 0, \quad c_{33}^0 = 1, \quad c_{11}^0 = 1/c_{22}^0 = (\lambda^0)^2; \qquad (29)$$

$$x_1 = 0: \quad v_1^0 = \overline{v}_1 \quad (v_2 = 0); \quad c_{12}^0 = 0, \quad c_{33}^0 = 1, \quad c_{11}^0 + c_{22}^0 + 1 = (\lambda^0)^2 + 2/\lambda^0. \qquad (29^*)$$

Here $\lambda_k^0$ are the components of the elastic stretching tensor for each mode at $x_1 = 0$ from the near field entrance jet flow, $\overline{v}_1$ is the rate average velocity in the die, and all the zero superscripts denote the variables at the entrance. It should be mentioned that unlike the initial conditions (29) for the plain die flow, the condition $c_{33}^0 = 1$ in (29*) resulting from the third equation in (23), makes impossible to use the continuity condition for elastic strains at $x_1 = 0$. It means that there is a small restructuring region near the die entrance cross-section. Using now in (29*) the incompressibility condition, still applicable to the entrance cross-section, and the condition $c_{12}^0 = 0$, results in the relation $c_{22}^0 = 1/c_{11}^0$, exactly the same as in (29). Therefore we can rewrite the boundary conditions (29*) for flow in cylindrical die in the equivalent form:

$$x_1 = 0: \; v_1^0 = \overline{v}_1 \; (v_2 = 0); \; c_{12}^0 = 0, \; c_{33}^0 = 1, \; c_{11}^0 = 1/c_{22}^0 = J(\lambda^0) + \sqrt{J^2(\lambda^0) - 1};$$
$$J(\lambda^0) = 1/2[(\lambda^0)^2 + 2/\lambda^0]. \tag{29**}$$

The problem formulated by equations (23)-(27) along with boundary conditions (28,28*), (29,29**) treats the viscoelastic developing flow in the entrance region as a transitional flow between the near field entrance, elongation jet flow and the developed shear Poiseuille flow, asymptotically achieved as $x_1 \to \infty$. It means that the developing flow in the die reflects the memory of the reservoir elongation entrance flow.

Under given flow rate condition, $Q = const$, the total pressure drop $\Delta P_{tot}$ between two pressure transducers, $T_1$ and $T_2$ shown in Fig.1, is now calculated as:

$$\Delta P_{tot} = \Delta P_1 + \Delta P_2 . \tag{30}$$

Here $\Delta P_1$ is the pressure difference due to the steady shear flow in the reservoir without converging flow region and $\Delta P_2$ is due to the shear flow in the slit die.

Calculations of initial elastic stretches $\lambda_k^0$, which are established from the analysis of near field entrance flow, represent the first necessary step in our computations of developing die flow. The next numerical step employs a finite difference numerical procedure with using a matrix solver for finding shear rate in each cross-sectional grid. From the boundary conditions (28) and (29), the initial values of variables, $\{v_1^0, c_{12}^0, c_{22}^0\}$ at $x_1 = 0$ are considered as known, and the values $\{c_{12}, c_{22}, \Delta P\}$ at the first step $\Delta x_1$ are found using the initial values. The same procedure is applied for calculations of all variables in cross-sections along the die.

## 3. RESULTS AND DISCUSSION

All the calculations in this paper are performed for isothermal viscoelastic flows under a given value of flow rate, using the general CE's (1), (2) with the specification of scaling relaxation factor $b$ given in Table 1. It should be noted that LDPE Melt 1 needs a special modeling [24,25] for the function $b$ as shown in Table 1 and Eqs. (1)-(2). The following necessary information is needed for comparing model calculations with experimental data or direct numerical simulations. We firstly need the rheological characterization of the polymer system, which could provide us with the knowledge of the discrete relaxation spectrum $\{G_k, \theta_k\}$. We secondly need the experiments or direct computations that present at least some space distributions of stress and/or velocity fields. We thirdly need a precise description of the geometry with a sharp entrance transition, used in these research papers.

Additionally the contraction flows of a dilute polymer solution are not fit for our model. Although it is not difficult to incorporate the pure viscous (solvent) term in Eqs. (1)-(2), the problem is that the flow *De* number for these solutions is not high enough to apply our modeling. We also think that a high increase in *De* number will also highly increase the Reynolds number, involving significant inertia effects. Thus because of infeasibility to compare our model calculations with data or direct numerical simulations for other polymer systems, we consider the results of below comparative studies only as preliminary.

Before discussing the comparison between calculations and data we should remind once again that both the flow *De* number and contraction ratio are significant in modeling of viscoelastic contraction flows.

For the slit channel flow, we use the experimental data [7] with discrete relaxation spectrum (Maxwellian modes) for a polyisobutylene (Vistanex). For the circular channel contraction flows, the data for another type of polyisobytylene (LM-MH) were used. We obtained the Maxwellian modes shown in *Table3* employing the Pade-Laplace procedure [26] from the linear dynamic data and stress relaxation experiments using the disc-disc fixture in RMS 800 rheometer. It should also be noted that our CE describes well the uniaxial elongation data obtained using an elongation rheometer of Messner type. The MPT (Monsanto Processability Tester) instrument with circular channel geometry and piston speed control was also used in our experiments. Along with the values of material characteristics, the values of geometrical parameters noted in *Fig. 1* are displayed in *Tables* 2 and 3. *Figure* 2 sketches various flow regions presented in our modeling. The far field reservoir entrance and the die flow are described as Poiseuille type flows and the near field reservoir flow is approximated as elongation flow type. Supplemental picture in *Figure* 2 also sketches the jet with secondary circulatory shear flow.

*Figure* 3 illustrates the comparison between our calculations and experimental data [7] for Vistanex. These data were obtained at certain axial positions for the near field entrance jet flow. Here the filled circles indicate experimental data and solid lines stand for our calculations with jet without drag. *Figure* 3 is suitable for die 2 (see *Table 2*). The data for stress invariant were obtained in [7] from birefringence experiments using the stress-optical "law":

$$[(\sigma_{11} - \sigma_{22})^2 + 4\sigma_{12}^2]^{1/2} = \Delta n / 1.414e - 9$$

Here $\Delta n$ is the birefringence and $\sigma_{ij}$ are the stress components. The first three plots in *Fig.* 3 show that model calculations (lines) disagree with experimental data (symbols).

The dashed lines in *Fig.* 3 stands for our calculations of the near-field flow involving additional viscoelastic shear circulatory flow. *Figure 3* presents an example of experimental data for the die 2 and our calculations for a higher speed flow with the higher, 7:1, contraction ratio. Here the flows in the reservoir and die regions have *De* number equal ( $De = \bar{\theta} \cdot \dot{\gamma}_{App} = 6.35 \cdot 3Q/2L \cdot L_R^2$ ) to 2.02 and 93.16, respectively. The experimental results in the first two plots of *Fig.* 3 show incomplete jet profiles with their initial low local *De* number. As the flow approaching to the die, the local *De* number increases, whereas the jet flow profile is building up. Once the jet profile occurs, the model predicts well enough the flow details for the rest of the region. Therefore we expect that for larger *De* numbers, the model assumption of shortening the transient length ($\delta \to 0$) between the far and near field reservoir regions will be more suitable. Having only few experimental points for flow in die 1, we cannot say that the additional shear drag in the near field entrance could significantly improve the stress distribution. But it makes much more improvement in our calculations of flow in the die 2 (*Fig.* 3.) as compared to the experimental data.

*Figure. 4*A presents the comparison between experimental data and our computations for the pressure difference between two transducers shown in *Fig.*1. Here we used Eq. (30) with certain flow rate values. According to Eq. (30), the near field reservoir region does not directly contribute in the total pressure. However, it indirectly affects the pressure drop in the developing die flow via the initial values of elastic strains. Therefore, as far as the model describes fairly the developing flow region with reasonable initial values at the die entrance, we expect a good agreement of our calculations with the data for plots of pressure drop versus flow rate. *Figure 4*A shows that both the types of above modeling, with and without account of jet's drag, coincide and describe well the data. Our computations also show that the extensional force with extra shear drag (dashed line) slightly increases the pressure drop ($\Delta P_1$) in the reservoir region as compared to the pure jet approach (solid line). It happens because of shortening the length of extended jet. However, the calculated pressures in the die ($\Delta P_2$) with the initial conditions, obtained by both model calculations, have almost the same values because $\Delta P_2 \gg \Delta P_1$. It means that both model types estimate quite successfully the initial values of elastic strains for the die flow. Also, as mentioned earlier, the increase in the contraction ratio and making the flow faster causes the increase in extensional force ($\Delta F_{ext}$). This effect is demonstrated in *Figure 4*B where one can clearly see the difference between the jet flow profiles without (solid line) and with the drag from the circulatory flow dashed line).

We now consider the comparison of our calculations with experimental data for another polyisobutylene (LM-MH). *Figure 5* (A-D) demonstrates basic rheological properties for this material. The data in Figs.5A,B,C were obtained using the disc-disc fixture of RMS 800 instrument, and the data in Fig.5.D, using an elongation rheometer of the Messner type. Employing the Pade-Laplace method [26], we obtained the value of parameters in discrete relaxation spectrum from the dynamic (A) and the stress relaxation (B) experiments in linear region. From steady simple shear experiments with different temperatures (Fig.5C), we found the activation energy using the Arrhenius equation, and confirmed the time-temperature superposition. Using in nonlinear region the multi-mode constitutive approach (1)-(3) with these Maxwell modes, we described experimental results in Fig. 6, including the uniaxial elongation (Fig.5D).

In the contraction flow with given circular capillary geometry shown in *Table 3*, (see also Fig.1) the contraction ratio is about 35, with the *De* number in reservoir ($De = 1.7 \cdot 4Q/\pi(R_R)^3$) 7.3e-3 and in die ($De = 1.7 \cdot 4Q/\pi(R_D)^3$) 304 for the highest flow rate 6.041e-8 (m$^3$/s). Because of the high difference between both the contraction ratios and *De* numbers, this situation is definitely favorable for exposing the effect of jet flow in the near field reservoir entrance region, as compared to the slit channel case. Figure 6 shows very good agreement with experimental data for the die flow within the range of *De* numbers from 6 to over 300.

*Figures* 7A,B demonstrates the centerline velocity and pressure profiles along the circular channel at *De* number about 250 using the jet flow models with and without drag, respectively. Because there is a big difference between the lengths of the two modeled reservoir jets, we changed the scale for negative (reservoir) values of longitudinal variable $x_1$ in *Fig.* 7A. Our calculations qualitatively agree with the experimental results obtained by Paskhin [1] who demonstrated that the centerline velocity goes first through the maximum and finally stabilizes. Direct comparison with data [1] was impossible because Paskhin used a reservoir with entering cone and also did not report important material characteristics. The comparison between the *Figures 7* A and B show that the simple jet model predicts longer length of reservoir jet and shorter length of the developing die flow region than the more realistic model of the reservoir jet with drag. E.g. *Fig.*7A shows that the longer dimensionless length ($\approx 16$) of the developing flow (evaluated by the axial velocity profile) needed for the jet with drag flow model as compared to that ($\approx 5$) for the jet flow only. The reason for this difference is longer relaxation of higher values of initial elastic strains at the die entrance, which is characteristic for the model of jet flow with drag. One can also see that for both the models of calculations, the length of developing flow, evaluated by the pressure profile, is considerably less than that evaluated by the axial velocity profile. It indicates that the stabilization of pressure gradient does not completely describe the developing flow region.

The contraction flow in Ref. [27] was investigated using LDPE melt (LDPE Lupolen 1810 H from BASF Ludwigshafen; Charge number 912 133 036). According to the experimental descriptions, the experiments were performed using 10:1 (reservoir/die) slit contraction controlled by the constant plunger speeds at a temperature of 150°C. At this temperature, the density of the material is $\rho = 778$ Kg/m$^3$ and the zero-shear-rate viscosity is approximately $\eta_o = 65$ KPa·s. The average axial velocities in the die were the ranged from 4.6 mm/s to 49.3 mm/s. No flow instability was observed. In addition, creeping flow conditions existed during all experiments. Laser Doppler velocimetry (LDV) was used to measure three-dimensional velocity field. The detail material characterization and 10:1 slit dimension are indicated in Table 1 and 4.
Here, instead directly employing their 14 Maxwellian modes, we used 8 modes from Laun [28] for characterizing LDPE. Preferably we try to use Pade-Laplace method [26], well-posed computer program to extract Maxwellian mode from the experimental data in linear region, but it could not happen due to no stress relaxation data available. Nevertheless using 8 modes by Laun and carefully chosen non-dimensional parameter shown in Table 1, we were able to describe all basic experiments as shown Fig. 8. Once again, there were two kinds of flow types used in our modeling, shearing and extensional, through all flow characteristic regions. Therefore the preliminary characterization of LDPE is expected working well for the model when it agrees reasonably as Fig. 8.
Upon our selection of Maxwellian modes, the average material characterization time is $\bar{\theta} = \sum_k G_k \theta_k^2 / \sum_k G_k \theta_k = 58.7s$ and the *De* numbers in the slit die are computed from $De = \bar{\theta} \cdot \dot{\gamma}_{App} = \bar{\theta} \cdot 6Q/(4LL_D^2)$ where $\dot{\gamma}_{App}$ is apparent shear arte in slit die, $Q$ is flow rate, $L$ is width of die and $L_D$ is the half thickness of die. Due to not clear indication of experimental flow rates (or *De* number) in the literature data [27], the flow rates are recalculated from the gap-wise axial velocity profile, which values are 4.00e-7, 1.02e-6 and 2.25e-6 m$^3$/s from the low *De* number.

*Figures 9* illustrates our model predictions and experiments for centerline axial velocity at 3 different *De* number flows. Figure 9A is computed using "Jet Approach" and Fig. 9B is done using "Jet Flow with Drag from Secondary Viscoelastic Flows" in the above section. Comparing Fig. 9A and B, the most significant difference in model prediction is reservoir converging length. For the same value of the given flow rate condition, the jet approach alone shows much longer converging length in reservoir than the jet with drag approach. It is because the extra force contribution by the secondary drag made the core jet shorten its converging path. In another words, the more force acting on the extensional flow, the less consuming time/path to reach from one cross-section area to another under the given condition.

Here with "jet approach" model, the near field flow predictions do not work well in Fig. 9A, because the demonstrated *De* numbers of LDPE may not be high enough. However, with "jet with drag", they show good agreements as Fig. 9B. Interestingly enough, with both model, the die flow predictions match well to the experimental data. Even if the more detail investigation of the die developing flow needs to be followed, the jet approach approximates the deformation profile at the die entrance roughly fair for the centerline velocities data.

On the comparison with the original simulations, presented in Fig.9C, our models make better results in the die entrance and developing region flow. The both figured the axial length about 10~20 in $x_1/L_D$ where the land region achieved. However, The original simulations under-estimate the die developing behaviors, even with higher values of experimental flow rate. Also it seems that the deviation of their simulation to experimental data increases for the higher *De* number flows.

*Figure 10* is showing the velocity profiles at $x_1/L_D = 40$ for each *De* numbers. The shown computations (lines) are obtained using the "jet with drag" model. Unfortunately the measurements are only available about at the land region, which may achieve by the steady state calculations. Therefore the detail developing behavior of die flow, the most interesting in the contraction flow, could not seen with Fig. 10. Nevertheless the all predictions, including jet approach models and the original literature simulation, agree as well as it shows with the experimental data.

## 4. CONCLUSIONS

Many published literature data do not provide necessary information for our modeling. Therefore we were unable to compare our model calculations with many direct calculations and/or experimental data. Nevertheless, the comparison of our model calculations with available literature results demonstrated that our model captures the most essential features of contraction flow with considerably less computational effort than direct computations. Thus we think that the main objective of our modeling, evaluating the entrance reservoir contraction flow and its effect on high *De* number polymer flow in die, is fulfilled.

There are remarkable advantages in our model predictions, such as better agreement with experimental data at the higher De numbers (high-speed processing) and higher contraction ratios, which are the well known as troublesome conditions for existing direct computational methods. Additionally (i) the model employs a complicated but very realistic rheological model in a geometrically simplified manner, and (ii) the model operates with well-defined rheological parameters and does not use any adjustable parameters.

Since the model approximately traces entire deformation history of the flow, it is possible to apply it for calculating swelling in short dies. This analysis will be demonstrated in the near future. Also computational analyses of non-isothermal contraction flows are relatively easy to carry out using the presented model. An example of such an analysis will be followed.

Other types of geometric and physical complexities, such as non-symmetrical entrance flows as a polymer wall slip, can also be treated within the presented model.

## APPENDIX A

**Matching Conditions for Different Types of Steady Channel Viscoelastic Flows**

We start from the exact local balance of the mechanical energy:

$$\frac{\partial}{\partial t}(K+W) + \frac{\partial}{\partial x_i}\left[v_i(K+W)\right] + D = \frac{\partial}{\partial x_i}(v_j \sigma_{ij}). \tag{A1}$$

Here $K = \rho|\underline{v}|^2$ and $W$ are the local kinetic and elastic energies, $D$ is the dissipation, $\underline{v}$ and $\underline{\underline{\sigma}}$ are the velocity and total stress. Eq.(A1) immediately follows from the momentum balance and general viscoelastic constitutive equations. Equation (A1) can be fractioned in the case of constitutive equation (1) in independent energy balances for each $k^{th}$ nonlinear relaxation modes:

$$\frac{\partial}{\partial t}(K+W_k) + \frac{\partial}{\partial x_i}\left[v_i(K+W_k)\right] + D_k = \frac{\partial}{\partial x_i}(v_j \sigma_{ij,k}), \tag{A2}$$

where $W_k$, $D_k$ and $\sigma_{ij,k}$ are respectively the elastic potential, dissipation, and the full stress tensor in the $k^{th}$ relaxation mode.

Consider now a composite steady flow of an incompressible viscoelastic liquid in a cylindrical tube whose arbitrary cross section is represented by a domain $\Omega$. Then integrating Eq.(A2) over the tube cross-section with the use of non-slip condition at the wall yields:

$$\frac{d}{dx_1}\int_\Omega v(K+W_k)dA = \frac{d}{dx_1}\int_\Omega v\sigma_{11,k}dA - \int_\Omega D_k dA. \tag{A3}$$

Here $v \equiv v_1$ is the longitudinal component of velocity. It should be mentioned that in any steady flow, the right-hand side of Eq.(A3) vanishes.

Let us now assume in the region $x_1 < x_0$ the flow is of a certain type, while in the region, $x_1 > x_0 + \delta$, it is of another type, with a transient flow in the region $x_0 < x_1 < x_0 + \delta$. Here $x_1$ is the longitudinal axis in flow region. If outside the transition region the flows could be approximately considered as steady, one can neglect the right-hand side in eq.(A3) and also neglect the contribution of the kinetic energy in the energetics of transition process. Then integrating (A3) over the transition area and neglecting the length of the transient zone $(\delta \to 0)$ will finally result in the jump-like continuity condition for the "free energy flux" in any $k^{\text{th}}$ relaxation mode:

$$\int_\Omega v W_k ds \bigg|_{x_0-0} = \int_\Omega v W_k ds \bigg|_{x_0+0}. \tag{A4}$$

The condition of the type (A4) has been used in paper [29] for evaluation of extrudate swell.

## APPENDIX B

### Evaluation of Stress Singularity in the Secondary Flow near the Upstream Corner

To simplify the evaluation of this singularity we consider in the expressions (20,20*) the drag from the circulatory Newtonian flow, and express the shear stress at the jet surface as:

$$\sigma_{12,surf} = -\frac{4\eta \bar{v}_1 / L_R}{(1-\xi)} \qquad (\xi = L_{Fx}/L_R) \tag{B1}$$

$$\sigma_{12,surf}^* = \frac{\eta \bar{v}_1}{R_R \xi^*} \left[ \frac{(1-\xi^{*2})(1-3\xi^{*2}) - 4\xi^{*4}\ln(\xi^*)}{(1-\xi^{*2})[(1-\xi^{*2}) + (1+\xi^{*2})\ln(\xi^*)]} \right] \qquad (\xi^* = R_{Fx}/R_R) \tag{B1*}$$

Here $\sigma_{12,surf}$ is the shear stress at the jet surface, $\eta$ is Newtonian viscosity, $\bar{v}_1$ is averaged axial velocity at $x_1 = -l$, $L_{Ex}$ ($R_{Ex}$) is half thickness (radius) of jet at $x_1 = -l + dx_1$, $L_R$ ($R_R$) is the half thickness (radius) of slit (circular) reservoir. Inserting (B1, B1*) into (19, 19*) and rearranging the formulae with the aid of (9) and (11), yields:

$$-\sum_k G_k (\lambda_k^2 + \frac{3}{\lambda_k^2})\frac{d\xi^*}{dx_1} = \frac{4\eta\bar{v}_1}{L_R^2(1-\xi^*)} + \sum_k \frac{G_k}{2\theta_k \bar{v}_1}(\lambda_k^4 - \frac{1}{\lambda_k^4}) \quad (\xi \to 1) \tag{B2}$$

$$-\sum_k G_k (\lambda_k^2 + \frac{2}{\lambda_k})\frac{d\xi^*}{dx_1} = \frac{\eta\bar{v}_1}{R_R^2(1-\xi^*)} + \sum_k \frac{G_k}{12\theta_k \bar{v}_1}(2\lambda_k + \frac{1}{\lambda_k^2})(\lambda_k^2 - \frac{1}{\lambda_k})(\lambda_k +1) \quad (\xi^* \to 1)$$

$$\tag{B2*}$$

Integrating Eqs.(20,20*) in the limits $\xi \to 1, \xi^* \to 1$ yields:

$$\xi \approx 1 - \sqrt{\alpha \Delta x_1^0}; \qquad \alpha = 8\eta \bar{v}_1 \left( L_R^2 \sum_k G_k (\lambda_k^2 + 3\lambda_k^{-2}) \bigg|_{\lambda_k = \lambda_k^l} \right)^{-1} \quad (\alpha \Delta x_1^0 << 1) \tag{B3}$$

$$\xi^* \approx 1 - \sqrt{\alpha^* \Delta x_1^0}; \quad \alpha^* = 2\eta \bar{v}_1 \left( R_R^2 \sum_k G_k (\lambda_k^2 + 2\lambda_k^{-1}) \bigg|_{\lambda_k = \lambda_k^l} \right)^{-1} \qquad (\alpha^* \Delta x_1^0 \ll 1) \qquad (B3^*)$$

Here $\lambda_k^l$ are the initial stretching values at the point $x_1 = -l$. Once the initial gap distance of secondary shear flow is determined from (B3, B3$^*$) for a small numerical step increase $\Delta x_1^0$, the iterative numerical method described in the main text, can be implemented.

The numerical procedure has been performed using a root finding subroutine for the unknown gap distance at $x_1 = -l + \Delta x_1^0$. And the shear stress at the jet surface $x_1 = -l + \Delta x_1^0$ has been computed from Eq. (B1, B1*) with the calculated gap distance value, $\xi$ (or $\xi^*$).

**Table 1.** Dimensionless parameters of the testing materials by Eq. (1) and (2)

| Polymer Melts | Dimensionless Parameters |
|---|---|
| Polyisobutylene (Vistanex) Isayev et al. [7] | $n = 0.1$ and $b(I_1, I_2; m) = 1$ |
| Polyisobutylene (LM-MH) | $n = 0.1$ and $b(I_1, I_2; m) = 1$ |
| LDPE (LDPE Lupolen 1810 H from BASF Ludwigshafen; Charge number 912 133 036) [27] | $n = 0.03$ and $b(I_1, I_2; m) = Exp[-\beta\sqrt{(I-3)}] + \dfrac{sinh[\nu(I-3)]}{[\nu(I-3)+1]}$ where $\beta = 0.35$, $\nu = 0.002$ |

**Table 2.** Material Parameters and Experimental Conditions for the Conversing and Flow by Fig.1-A.

| Polymer Melts | Parameters at 300K | | Conditions |
|---|---|---|---|
| | $\theta_k$ (sec) | $G_k$ (KPa) | |
| Polyisobutylene (Vistanex) Isayev *et al.* [7] | 7.025 | 4.55 | *Geometry* (cm) $L = 2.0$, $L_z = 3.5$, $l_a = 4.0$, $l_b = 2.02$, $l_c = 0.95$ |
| | 1.553 | 11.59 | Die 1: $2L_R = 0.8$, $2L_D = 0.4$ |
| | 0.182 | 86.81 | Die 2 $2L_R = 0.822$, $2L_D = 0.121$ |

**Table 3.** Material Parameters and Experimental Conditions for the high *De* number Capillary Flow by Fig.1-B.

| Polymer Melts | Parameters at 323K | | Conditions |
|---|---|---|---|
| | $\theta_k$ (sec) | $G_k$, (KPa) | |
| Polyisobutylene (LM-MH) | .006 | 144.72 | *Geometry* (cm) (Mosanto Processability Tester) Reservoir Radius ($R_R$) = 2.6162 Die Radius ($R_D$) = 0.07544 Die Length ($L_z$) = 3.0176 |
| | .046 | 47.38 | |
| | .204 | 19.63 | |
| | .815 | 7.517 | |
| | 4.091 | 1.656 | |

**Table 4.** Material Parameters and Experimental Conditions for the Conversing and Flow by Fig.1-A.

| Polymer Melts | Parameters at 423K [28] | | Conditions |
|---|---|---|---|
| | $\theta_k$ (sec) | Gk, (KPa) | |
| LDPE (LDPE Lupolen 1810 H from BASF Ludwigshafen; Charge number 912 133 036) [27] | 1000 | 0.001 | Geometry (cm) |
| | 100 | 0.18 | |
| | 10 | 1.89 | L = 2.0 |
| | 1 | 9.8 | Reservoir: |
| | 0.1 | 26.7 | $2L_R$ = 2.0, $L_Z$ = -10.0 |
| | 0.01 | 58.6 | Die: |
| | 0.001 | 94.8 | $2L_D$ = 0.2, $L_Z$ = 5.0 |
| | 0.0001 | 129 | |

**Figure Captions**

**Fig.1** Sketches of the slit (A) and circular (B) rheometers:

1 – reservoir, 2- die, *T - the pressure transducer locations in the slit/circular rheometers.

**Fig.2** Schematic classification of contraction flow regions.

**Fig.3** Gap-wise profiles of stress invariant for PIB (Vistanex) contraction flow in the die 2 geometry at various axial locations at $27^{o}C$ , with the flow rate, $Q = 7.16e-8$ ($m^3$/s): filled circles – experiments; solid lines – jet calculations without drag; dashed lines – jet calculations with drag.

**Fig.4** Contraction flow of PIB (Vistanex) at 27 $^{o}C$: solid lines – jet calculations without drag; dashed lines - jet calculations with drag.
**A.** The plots of pressure differences between two mounted transducers (*$T_1$ and *$T_2$ in Fig. 1 slit rheometer) versus flow rate $Q$: o- experimental data for die 1, ☐ - experimental data for the die 2.
**B:** Non-dimensional plots of the jet profiles, calculated with and without drag, versus axial location in the reservoir. Flow in the die 2 at flow rate $Q = 7.16e-8 m^3$/s s

**Fig.5** Basic rheological experimental data and their modeling for PIB (LM-MH) for:
**A -** dynamic test at 50 $^{o}C$,
**B -** stress relaxation test at 50 $^{o}C$,
**C -** temperature dependence of Newtonian viscosity,
**D -** uniaxial extension test at 18 $^{o}C$.
Various symbols in the Figures indicate experimental data, lines denote calculations.

**Fig.6** Isothermal circular capillary flow for PIB (LM-MH) at 50 $^{o}C$ in Monsanto Processability Tester (geometrical description is given in Table 3):
o- experimental data; solid lines – jet calculations without drag; dashed lines- with drag.
**A:** Plots of apparent viscosity versus apparent shear rate.
**B:** Plots of transducer *$T$ pressure (Fig 1B) versus flow rate.

**Fig.7** The flow in the circular rheometer for PIB (LM-MH) at 50 $^{o}C$ under flow rate, $Q = 5.0e-8$ ($m^3$/s): solid lines – jet calculations without drag; dashed lines- jet calculations with drag.
**A:** Centerline axial velocity profiles in the reservoir and die regions.
**B:** Pressure profile along the die region

**Fig.8** Basic rheological experimental data and the CE predictions for LDPE:
**A –** steady simple shear viscosity and the first normal stress difference $N_1$ at 150 $^{o}C$,
**B –** dynamic test at 150 $^{o}C$,

**C -** uniaxial extension test at 150 $^{\circ}$C: various symbols stand for experimental data and lines for calculations.

**Fig.9** Experimental data for LDPE and calculations for the centerline axial velocity under various given flow rates at 150 $^{\circ}$C; the origin of $x_1$ locates at the die entrance.

**A –** Computations using the jet approach,

**B –** Computations using the jet approach with drag,

**C –** Direct computation and experimental data [27]. Symbols stand for experimental data and lines for calculations.

**Fig.10** Gap-wise axial velocity profile with various flow rates. Symbols stand for experimental data and lines for calculations.

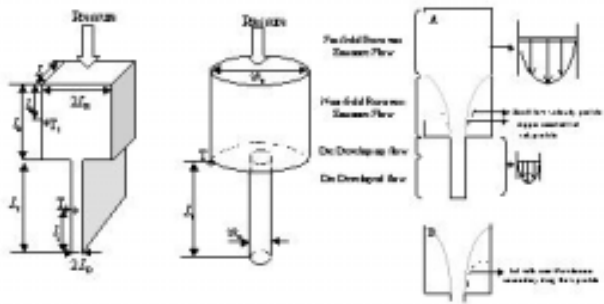

Fig.1 Fig.2

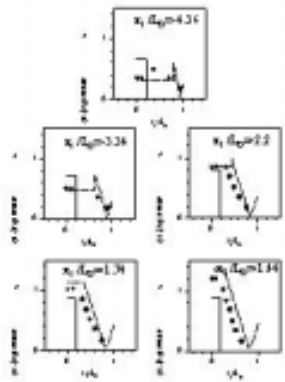
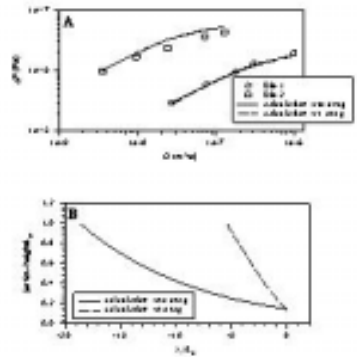

Fig.3 Fig.4

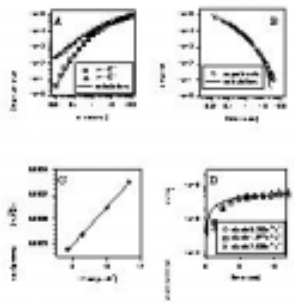
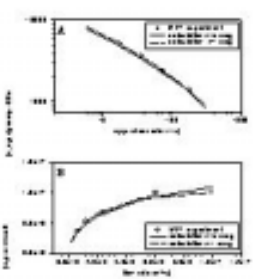

Fig.5 Fig.6

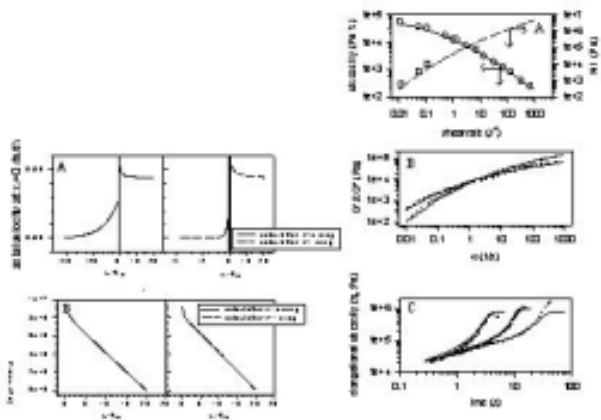

Fig.7    Fig.8

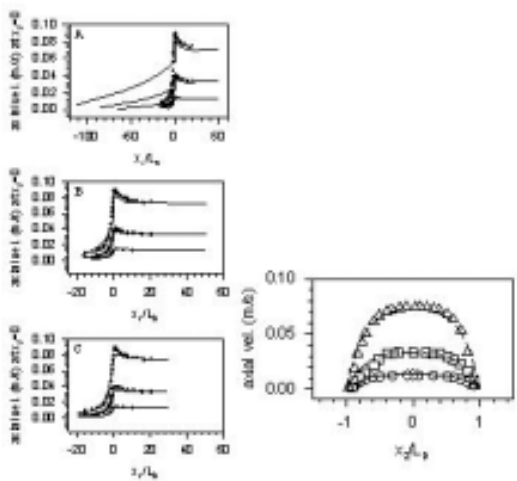

Fig.9    Fig.10